\documentclass[aps,prd,preprintnumbers,superscriptaddress,nofootinbib,showpacs]{revtex4}
\usepackage[dvips]{graphicx}
\usepackage{bm,latexsym,amsmath,amssymb,amsfonts,mathrsfs}
\usepackage{color}
\input{colordvi.tex}
\newcommand*{\D}{{\rm d}}
\newcommand*{\cs}{c_{\rm s}}
\newcommand*{\cscr}{c_{{\rm s}*}}

\newcommand*{\mpl}{M_{\rm Pl}}
\begin{document}

\title{Primordial non-Gaussianity from G-inflation}

\author{Tsutomu~Kobayashi}
\email[Email: ]{tsutomu"at"tap.scphys.kyoto-u.ac.jp}
\affiliation{Research Center for the Early Universe (RESCEU), Graduate School of Science,
The University of Tokyo, Tokyo 113-0033, Japan\footnote{Present address: Hakubi Center,
Kyoto University, Kyoto 606-8302, Japan
and
Department of Physics, Kyoto University, Kyoto 606-8502, Japan}}

\author{Masahide~Yamaguchi}
\email[Email: ]{gucci"at"phys.titech.ac.jp}
\affiliation{Department of Physics, Tokyo Institute of Technology, Tokyo 152-8551, Japan}

\author{Jun'ichi~Yokoyama}
\email[Email: ]{yokoyama"at"resceu.s.u-tokyo.ac.jp}
\affiliation{Research Center for the Early Universe (RESCEU), Graduate School of Science,
The University of Tokyo, Tokyo 113-0033, Japan}
\affiliation{Institute for the Physics and Mathematics of the Universe(IPMU),
The University of Tokyo, Kashiwa, Chiba, 277-8568, Japan}

\begin{abstract}
We present a comprehensive study of primordial fluctuations generated from G-inflation, in which the inflaton Lagrangian
is of the form $K(\phi, X)-G(\phi, X)\Box\phi$ with $X=-(\partial\phi)^2/2$.
The Lagrangian still gives rise to second-order gravitational and scalar field equations,
and thus offers a more generic class of single-field inflation than ever studied,
with a richer phenomenology.
We compute the power spectrum and the bispectrum, and clarify
how the non-Gaussian amplitude depends upon parameters such as the sound speed.
In so doing we try to keep as great generality as possible, allowing for non slow-roll
and deviation from the exact scale-invariance.
\end{abstract}

\preprint{RESCEU-3/11}
\maketitle

\section{Introduction}

Cosmological inflation~\cite{inflation} is now a widely accepted paradigm explaining the
flatness, homogeneity, and isotropy of the observed Universe.
In the most common scenario, inflation occurs when
the inflaton, a scalar field driving the accelerated expansion, rolls down
a nearly flat potential slowly.
During this slow-roll stage
fluctuations in the inflaton field are generated quantum-mechanically
and stretched outside the Hubble horizon, which eventually reenter
the Hubble radius in a later epoch to be
a seed for the large-scale structure of the Universe.
The detailed shape of the potential can be probed
by observing the power spectrum of fluctuations 
in terms of the cosmic microwave background (CMB) anisotropies~\cite{wmap}.
As to theoretical approaches, much effort has been made to determine
the inflaton potential in the particle physics context.
However, single-field inflation with a canonical kinetic term and
a nearly flat potential is not the only option to induce
the accelerated expansion and to produce
almost scale-invariant perturbations with an appropriate amplitude.
Liberating inflation models from the standard assumption,
one may consider a variety of interesting scenarios:
multiple scalar fields might participate the inflationary dynamics,
the kinetic term of the inflaton(s) might be non-canonical~\cite{kinflation}, and
a scalar field other than the inflaton might be responsible for the density
perturbation~\cite{curvaton}.
From a high-energy physics point of view, supersymmetric theories naturally provide many scalar
fields with flat potentials~\cite{susyinf}, and the Dirac-Born-Infeld (DBI)-type non-canonical
kinetic term naturally arises from D3-brane motion in a warped compactification~\cite{DBI}.

Different inflationary scenarios can be distinguished
by future and on-going experiments such as Planck~\cite{PLANCK}, aiming to obtain
better constraints on
the amount of non-Gaussianities in the primordial curvature perturbations
as well as on the spectral index $n_s$, its running,
and the tensor-to-scalar ratio $r$.
The standard canonical slow-roll inflation models
produce negligible non-Gaussianity~\cite{Malda},
while exotic inflationary scenarios are expected to predict
measurable non-Gaussian signals.
In the context of single-field inflation,
non-Gaussian perturbations have been computed for
the Lagrangian of the form~\cite{LS, Kachru}
\begin{eqnarray}
{\cal L}_\phi=K(\phi, X), \label{Lag-K}
\end{eqnarray}
where $\phi$ is the inflaton and $X:=-\partial_\mu\phi\partial^\mu\phi/2$.
This class of models yields a sound speed $\cs$
different from the speed of light in general,
and large non-Gaussianity is generated for $\cs \ll 1$.
A significant non-Gaussian signal together with the confirmation of
the consistency relation $r=-8\cs n_T$, where $n_T$ is
the spectral index of primordial tensor perturbations,
is a smoking gun of the inflaton Lagrangian~(\ref{Lag-K}).

In this paper, we consider a more general Lagrangian~\cite{vikman, GI}
\begin{eqnarray}
{\cal L}_\phi = K(\phi, X)-G(\phi, X)\Box\phi,\label{Galileon-Lagrangian}
\end{eqnarray}
where $K$ and $G$ are some generic functions of the inflaton
$\phi$ and $X$.
The new term $G(\phi, X)\Box\phi$ in the Lagrangian~(\ref{Galileon-Lagrangian})
is inspired by the Galileon interaction~\cite{G1, G2} and reduces to the one having
the Galilean shift symmetry, $\partial_\mu\phi\to\partial_\mu\phi+b_\mu$,
in the Minkowski background in the case $G\propto X$.
One of the most important properties of the Galileon Lagrangian is that
the field equations do not contain derivatives higher than two.
The interaction $G(\phi, X)\Box\phi$ is a generalization of
the Galileon term $X\Box\phi$ while maintaining the second-order property.
In this sense, the Lagrangian~(\ref{Galileon-Lagrangian}) defines
a more generic class of single-field inflation than ever studied.
Here, the Galilean shift symmetry is abandoned in exchange for generality,
but one should note that the symmetry does not make sense already upon covariantization
for any interaction that is Galilean invariant in the flat background.\footnote{Concerning
this point, one may worry about the naturalness of G-inflation models discussed in the
present paper because there is no symmetry to protect the Lagrangian. However, it should be noted
that symmetry, if present, must be broken at least to end inflation. We therefore will not
provide a symmetry-based argument
but rather take a phenomenological approach, assuming that some UV complete theory
would give the (in some sense fine-tuned) Lagrangian that leads to second-order
field equations.}
(The name ``Galileon'' is therefore no longer appropriate
when covariantized.)
Cosmological applications of the Galileon
interaction can be found in~\cite{gde}
with emphasis on dark energy and modified gravity.
Primordial inflation based on the generic Lagrangian~(\ref{Galileon-Lagrangian})
was first proposed very recently by~\cite{GI, HGI}, and is dubbed {\em G-inflation}.
Almost simultaneously the same Lagrangian was used to explain
the late-time cosmic acceleration rather than the primordial one~\cite{vikman, Kimura}.
In~\cite{ginf1, ginf2}
the effective-field-theory approach~\cite{eft} was employed
to see the consequences of imposing
the approximate Galilean shift symmetry
on the Lagrangian of primordial perturbations.
Interestingly, the scalar field theory with the $G\Box\phi$ term
can violate the null energy condition stably. This fact motivates the authors of
Refs.~\cite{NV, Genesis} to propose a radical scenario of the earliest Universe
alternative to inflation.
Some specific form of the above type of interaction arises
from a probe brane action in higher dimensions~\cite{dbigalileon} and from
the Kaluza-Klein reduction
of Lovelock gravity~\cite{kkreduction, bstring}.
A supersymmetric completion of Galileons is explored in~\cite{susy}.

The purpose of the present paper is to understand the nature of
cosmological perturbations generated from G-inflation.  We rederive the
power spectrum and the tilt of the spectrum without assuming slow-roll,
clarifying how the (approximate) scale-invariance is achieved in
G-inflation.  We then calculate the cubic action for the curvature
perturbation and evaluate the full non-Gaussian amplitude, again without
assuming slow-roll and the exact scale-invariance.  Throughout the paper
we try to make our formulas as general as possible, which we hope
maximizes the usefulness of the results. Recently, non-Gaussianity
from G-inflation was calculated neglecting a number of terms working in
the de Sitter limit \cite{MK} and in the slow-roll limit~\cite{DeFelice:2011zh}.
See also a recent work by Naruko and Sasaki,
in which the superhorizon evolution of the nonlinear curvature
perturbation from G-inflation is addressed~\cite{naruko}.

This paper is organized as follows.
In the next section we review the basic properties of G-inflation
and derive the power spectrum of the curvature perturbation.
In Sec.~III we compute the cubic action for the curvature perturbation
to evaluate the three-point function in G-inflation.

\section{G-inflation}

We start with a brief review on the basics of G-inflation~\cite{GI, HGI}.
The scalar field Lagrangian for G-inflation is given by Eq.~(\ref{Galileon-Lagrangian}).
Assuming that $\phi$ is minimally coupled to gravity,
the total action we are going to study is
\begin{eqnarray}
S=\int\D^4x\sqrt{-g}\left[\frac{\mpl^2}{2}R+{\cal L}_\phi\right].\label{action}
\end{eqnarray}
In the following we will set $M_{\rm Pl} =1$.
The energy-momentum tensor $T_{\mu\nu}$ of the scalar
field is given by
\begin{eqnarray}
T_{\mu\nu}=K_{X}\nabla_\mu\phi\nabla_\nu\phi+K g_{\mu\nu}-2\nabla_{(\mu}G \nabla_{\nu)}\phi
+ g_{\mu\nu}\nabla_{\lambda }G\nabla^{\lambda }\phi
-G_X\Box\phi\nabla_\mu\phi\nabla_\nu\phi.\label{tmn}
\end{eqnarray}
Here and hereafter we use the notation $K_X$ for $\partial K/\partial X$ etc.
Varying the action with respect to $\phi$, we obtain the scalar field equation of motion,
\begin{eqnarray}
&&K_{X}\Box\phi-K_{XX}(\nabla_{\mu}\nabla_\nu\phi)(\nabla^\mu\phi\nabla^\nu\phi)
-2K_{\phi X}X+K_\phi
-2\left( G_\phi- G_{\phi X}X\right)\Box\phi
\nonumber\\&&
+G_X\left[
(\nabla_\mu\nabla_\nu\phi)( \nabla^\mu\nabla^\nu\phi)-(\Box\phi)^2+R_{\mu\nu}\nabla^\mu\phi\nabla^\nu\phi
\right]
+2G_{\phi X}(\nabla_{\mu}\nabla_\nu\phi)(\nabla^\mu\phi\nabla^\nu\phi)+2G_{\phi\phi}X
\nonumber\\&&
-G_{XX}\left(
\nabla^\mu\nabla^\lambda\phi -g^{\mu\lambda}\Box\phi
\right)\left(\nabla_\mu\nabla^\nu\phi\right)\nabla_\nu\phi\nabla_\lambda\phi
=0,\label{sfeom}
\end{eqnarray}
which is of course equivalent to the conservation equation $\nabla_\nu T_\mu^{\;\nu}=0$.
One verifies from Eqs.~(\ref{tmn}) and~(\ref{sfeom})
that the gravitational and scalar field equations are indeed of second order.

Higher order Galileon terms (with a $\phi$-dependent coefficient) such as
$f(\phi)X\left[2(\Box\phi)^2-2\nabla_\mu\nabla_\nu\phi\nabla^\mu\nabla^\nu\phi+RX\right]$
can be added to the scalar field Lagrangian while keeping the
field equations of second order.
Although the effect of such higher order Galileons might be interesting
in the context of primordial inflation, we leave the issue for future study and
concentrate on the Lagrangian of the form~(\ref{Galileon-Lagrangian}) in the present paper.

\subsection{The background equations}

Let us consider homogeneous and isotropic background:
\begin{eqnarray}
\D s^2=-\D t^2+a^2(t)\D\mathbf{x}^2,\quad \phi=\phi(t).
\end{eqnarray}
Although the energy-momentum tensor~(\ref{tmn}) cannot be recast in a perfect-fluid form
in general~\cite{vikman}, for the above cosmological ansatz
it takes the desirable form
$T_\mu^{\;\nu}=\text{diag}(-\rho, p, p, p)$ with
\begin{eqnarray}
\rho&=&2K_XX-K+3HG_X\dot\phi^3-2G_\phi X,
\\
p&=&K-2\left(G_\phi+G_X\ddot\phi \right)X.
\end{eqnarray}
The gravitational field equations are thus
\begin{eqnarray}
3H^2&=&\rho,
\\
-3H^2-2\dot H &=&p,
\end{eqnarray}
and the scalar field equation of motion is given by
\begin{eqnarray}
&& K_X \left( \ddot{\phi}+3H\dot{\phi} \right) + 2 K_{XX} X \ddot{\phi} +
   2K_{X\phi} X - K_{\phi}
-2\left( G_{\phi}-G_{X\phi}X \right)\left( \ddot{\phi}+3H\dot{\phi} \right) 
\nonumber\\&&
+ 6G_{X} \left[ \left( HX \right)\dot{}+3H^2 X \right]
- 4G_{X\phi} X \ddot{\phi} - 2G_{\phi\phi}X+6HG_{XX}X \dot{X}=0.
\end{eqnarray}

If, for example, $K$ is given by the standard, canonical kinetic term
with a potential, $K=X-V(\phi)$, one can consider an inflationary
scenario in which the energy density is dominated by the potential as in
the standard case, while the dynamics of the scalar field is modified by
the $G\Box\phi$ term, changing the potential that $\phi$ effectively
feels.  This is the scenario proposed in~\cite{HGI} and called
potential driven G-inflation. Another possible scenario is that
inflation is driven by $\phi$'s kinetic energy which is kept almost
constant with nontrivial functional form of $K$ and $G$.  In models with
the exact shift symmetry, $\phi\to\phi+c$, {\em i.e.}, $K=K(X)$ and
$G=G(X)$, it is easy to obtain an exactly de Sitter background
satisfying $H=$ const and $\dot\phi=$ const.  This may be regarded as a
generalization of k-inflation~\cite{kinflation}, and we call the class
of models kinematically driven G-inflation~\cite{GI}.  Deferring the
summary of these two specific classes of G-inflation to
Sec.~\ref{models}, we now move on to describe the general properties of
the power spectrum of primordial perturbations from G-inflation.

\subsection{Power spectrum}\label{1st-order}

In this section we derive a series of general formulas for linear cosmological
perturbations without assuming any specific form of $K$ and $G$.
We work in the unitary gauge, $\phi(t,\mathbf{x}) =\phi(t)$.\footnote{The unitary gauge
does not coincide with the comoving gauge, $\delta T_i^{\;0}=0$, in the case
of G-inflation~\cite{GI}. This fact stems from the imperfect-fluid nature of
the energy-momentum tensor~(\ref{tmn}).}
Using the remaining gauge degree of freedom the
linearly perturbed metric is taken to be
\begin{eqnarray}
\D s^2=-(1+2\alpha_1)\D t^2+2
a^2\partial_i\beta_1\D t\D x^i+a^2(1+2{\cal R})\D\mathbf{x}^2.
\label{1st-order-metric}
\end{eqnarray}
Expanding the action to second order in perturbations and then
varying with respect to $\alpha_1$ and $\beta_1$,
we obtain the following constraint equations:
\begin{eqnarray}
\dot{\cal R}&=&\Theta\alpha_1,\label{cst1}
\\
\frac{\partial^2}{a^2}\left({\cal R}+a^2\Theta\beta_1\right)&=&X{\cal G}\alpha_1,\label{cst2}
\end{eqnarray}
where $\partial^2:=\delta^{ij}\partial_i\partial_j$,
\begin{eqnarray}
\Theta&:=&H-\dot\phi XG_X,
\\
{\cal G}&:=&K_X+2XK_{XX}+6G_{X}H\dot\phi+6G_X^2X^2
-2\left(G_\phi+XG_{\phi X}\right)+6G_{XX}HX\dot\phi.
\end{eqnarray}
Substituting the constraints~(\ref{cst1}) and~(\ref{cst2}) to the action,
we arrive at
the quadratic action for ${\cal R}$~\cite{GI, vikman}:
\begin{eqnarray}
S_2=\int\D t\D^3x\, a^3\sigma\left[\frac{1}{\cs^2}\dot{\cal R}^2
-\frac{1}{a^2}(\partial{\cal R})^2\right],
\label{2nd-action}
\end{eqnarray}
where
\begin{eqnarray}
\cs^2&:=&\frac{{\cal F}}{{\cal G}},
\\
\sigma&:=&\frac{X{\cal F}}{\Theta^2},
\end{eqnarray}
and
\begin{eqnarray}
{\cal F}&:=&K_X+2G_X\left(\ddot\phi+2H\dot\phi\right)
-2G_X^2X^2
+2G_{XX}X\ddot\phi-2\left(G_\phi-XG_{\phi X}\right).
\end{eqnarray}
One can verify that setting $G(\phi, X)=0$
the quadratic action~(\ref{2nd-action}) reproduces
the expression obtained for k-inflation~\cite{per-k-inf}.
It is useful to notice that
$\sigma$ can also be expressed as
\begin{eqnarray}
\sigma =-\frac{\dot\Theta}{\Theta^2}+\frac{\dot\phi XG_X}{\Theta}.
\label{sigma-theta}
\end{eqnarray}

Let us define three parameters that characterize
the rate of change of three background quantities:
\begin{eqnarray}
\epsilon:=-\frac{\dot H}{H^2},
\quad
s:=\frac{\dot\cs}{H\cs},
\quad
\delta:=\frac{\dot\sigma}{H\sigma}.
\label{slrpar}
\end{eqnarray}
In this paper we assume that
\begin{eqnarray}
\frac{\dot\epsilon}{H\epsilon}\simeq 0,
\quad
\frac{\dot s}{Hs}\simeq 0,
\quad
\frac{\dot\delta}{H\delta}\simeq 0,
\end{eqnarray}
but
we do not neglect $\epsilon$, $s$, and $\delta$.
(In the next section, however, we will assume some stronger conditions to evaluate the bispectrum.)
It should be noted in particular that
$\sigma$ is not necessarily small,
in contrast to the usual (k-)inflation models
in which $\sigma$ is degenerate, {\em i.e.}, $\sigma =\epsilon<1$~\cite{per-k-inf}.
Even in the slow-roll limit we may have $\sigma\gtrsim 1$ in G-inflation.


Under the assumption that the parameters defined in~(\ref{slrpar})
are constant (but not necessarily very small),
it is straightforward to solve the equation of motion derived from the action~(\ref{2nd-action})
and compute the power spectrum of ${\cal R}$~\cite{GI}.
For this purpose it is convenient to define
a new time coordinate $y$ by $\D y = \cs \D t/a$~\cite{KP}.
In terms of $y$,
the scale factor, the sound speed, and $\sigma$ are written as
\begin{eqnarray}
a=\frac{c_{{\rm s}*}(y/y_*)^{-1/(1-\epsilon -s)}}{(-y_*)H_*(1-\epsilon -s)},
\quad
\cs =c_{{\rm s}*}(y/y_*)^{-s/(1-\epsilon-s)},
\quad
\sigma=\sigma_*(y/y_*)^{-\delta/(1-\epsilon-s)},
\end{eqnarray}
where the quantities with $*$ are those evaluated at some reference time $y=y_*$.
Using
a new variable $u:=\tilde z{\cal R}$ with $\tilde z:=a\sqrt{2\sigma/\cs}$,
the equation of motion can be written in the Fourier space as
\begin{eqnarray}
u_k''+\left(k^2-\frac{\tilde z''}{\tilde z}\right)u_k=0,\label{SMeq}
\end{eqnarray}
where the prime denotes differentiation with respect to $y$ and
we find
\begin{eqnarray}
\tilde z\propto (-y)^{1/2-q},\quad
\frac{\tilde z''}{\tilde z}=
\frac{q^2-1/4}{y^2},
\quad{\rm with}\quad
q:=\frac{3-\epsilon -2s +\delta}{2(1-\epsilon -s)}.
\end{eqnarray}
The normalized mode solution to Eq.~(\ref{SMeq})
corresponding to the Minkowski vacuum in the high frequency limit
is then given in terms of
the Hankel function by
\begin{eqnarray}
u_k=\frac{\sqrt{\pi}}{2}\sqrt{-y}H_{q}^{(1)}(-ky).\label{def-psi}
\end{eqnarray}
We thus write the operator ${\cal R}$ using the creation and annihilation modes as
\begin{eqnarray}
{\cal R}(\mathbf{k}, y) &=&\psi(k,y)\hat a_{\mathbf{k}}+\psi^*(-k, y)\hat a^\dagger_{-\mathbf{k}},
\label{Raa}
\\
\psi(k, y) &=& \frac{u_k(y)}{\tilde z},
\end{eqnarray}
with the commutation relation
$[\hat a_{\mathbf k},\hat a^\dagger_{{\mathbf k'}}]=(2\pi)^3\delta^{(3)}({\mathbf k}-{\mathbf k}')$.
This immediately leads to the power spectrum~\cite{GI},
\begin{eqnarray}
{\cal P}_{\cal R}=\frac{k^3}{2\pi^2}\left|\frac{u_k}{\tilde z}\right|^2
=2^{2q-3}\left|\frac{\Gamma(q)}{\Gamma(3/2)}\right|^2
\left.\frac{(1-\epsilon -s)^2}{4\pi^2}
\frac{H^2}{2\sigma\cs}\right|_{ky=-1}.
\label{Ps}
\end{eqnarray}

The scalar spectral index is found to be
\begin{eqnarray}
n_s-1=3-2q=-\frac{2\epsilon+s+\delta}{1-\epsilon-s}.
\end{eqnarray}
The above formula has been derived without assuming the smallness of $\epsilon, s$ and $\delta$,
though we have assumed that they are constant.
In this sense, the above expression is more general than that given in~\cite{GI, HGI, MK,DeFelice:2011zh}.
To ensure the scale invariance we require
$2\epsilon + s +\delta \simeq 0$.
However, this does not force {\em each} parameter
to be as small as ${\cal O}(n_s-1)$;
each can be large, $\epsilon, s, \delta\gg{\cal O}(n_s-1)$, but the three
may cancel each other out to produce an almost scale-invariant
spectrum. This possibility was first
pointed out by~\cite{KP} in the less generic context of DBI inflation,
for which $\sigma=\epsilon$ and consequently $\delta=0$.
We leave this interesting possibility open, and will complete the following calculation
without taking the slow-roll limit.
We would stress again that
even if we consider the slow-roll limit, $\sigma$ is not necessarily slow-roll suppressed.


Since the inflaton field is minimally coupled to gravity,
the nature of tensor perturbations is the same as the standard one
and is dependent only on the geometrical quantity $H=H(t)$.
In the slow-roll limit, $\epsilon=0$,
the tensor power spectrum is given by ${\cal P}_h= 8(H/2\pi)^2$.
The tensor-to-scalar ratio $r$ is thus given by
\begin{eqnarray}
r=16\sigma\cs,
\end{eqnarray}
where just for simplicity
the scalar power spectrum is evaluated also in the slow-roll limit,
$\epsilon=s=\delta =0$.

For later convenience we introduce the following quantity:
\begin{eqnarray}
\nu:=\frac{\dot\phi XG_X}{H},
\end{eqnarray}
or, equivalently, $\Theta=H(1-\nu)$.
From Eq.~(\ref{sigma-theta}) we obtain
\begin{eqnarray}
\sigma = \frac{\dot\nu}{H(1-\nu)^2}+\frac{\nu}{1-\nu}+\frac{\epsilon}{1-\nu}.
\label{sigma-nu-rel}
\end{eqnarray}
For $\epsilon=$ const, $s=$ const, and $\delta =$ const, the above
equation can be integrated to yield
\begin{eqnarray}
\frac{H}{\Theta}=
\frac{1}{1-\nu(y)} =\frac{1}{1+\epsilon}+\frac{\sigma(y)}{1+\epsilon+\delta}
+\left(\frac{1}{1-\nu_*}-\frac{1}{1+\epsilon}
-\frac{\sigma_*}{1+\epsilon+\delta}\right)(y/y_*)^{(1+\epsilon)/(1-\epsilon-s) }.
\end{eqnarray}
If we assume $\nu=$ const then we have $\sigma =$ const.
In this case the two quantities are related as
\begin{eqnarray}
\nu=\frac{\sigma -\epsilon }{1+\sigma }.
\end{eqnarray}
Note in passing that the opposite is not in general true:
for $\sigma=$ const Eq.~(\ref{sigma-nu-rel}) still admits time-dependent $\nu$.

\subsection{G-inflation examples}\label{models}

\subsubsection{Kinematically driven G-inflation}

Inflation can be driven by kinetic energy of $\phi$.
This possibility was explored in~\cite{GI}.
Let us consider for simplicity
the Lagrangian with exact shift symmetry $\phi\to\phi+c$, {\em i.e.},
\begin{eqnarray}
K=K(X),\quad G=G(X),
\end{eqnarray}
and look for an exact de Sitter background
satisfying $H=$ const and $\dot\phi=$ const.
It follows from the field equations that
\begin{eqnarray}
&&
3H^2=-K,\label{KG1}\\
&&
K_X+3G_XH\dot\phi =0.\label{KG2}
\end{eqnarray}
For this background we have
\begin{eqnarray}
{\cal F} &=& -\frac{K}{3X}\nu(1-\nu),
\\
{\cal G} &=&-\frac{K}{X}\nu\left(1+\nu-2\frac{XK_{XX}}{K_X}+2\frac{XG_{XX}}{G_X}\right),
\\
\sigma &=&\frac{\nu}{1-\nu},
\end{eqnarray}
where $\nu=\dot\phi XG_X/H=XK_X/K=$ const.
In evaluating the above equations
we used the background equations~(\ref{KG1}) and~(\ref{KG2}).

The concrete toy model presented in~\cite{GI} is given by
\begin{eqnarray}
K=-X+\frac{X^2}{2M^3\mu},\quad
G=\frac{X}{M^3},\label{gimodel}
\end{eqnarray}
where $M$ and $\mu$ are parameters.
In this case, $\cs$ and $\sigma$ can be expressed in terms of $\mu$.
It turns out that the tensor-to-scalar ratio $r=16\sigma\cs = 16\sigma \cs(\sigma)$
is an increasing function of $\sigma$, and $\sigma\simeq\nu\ll 1$
is required in order for $r$
not to exceed the observationally allowed value.
Explicitly, one finds
$r\simeq (8/\sqrt{3})\sigma^{3/2}\simeq (16\sqrt{6}/3)(\sqrt{3}\mu)^{3/2}$~\cite{GI}.

Note, however, that $\nu\ll 1$ is not necessary
to get a stable, prolonged de Sitter phase.
As already emphasized above,
$\sigma\gtrsim 1$ is made possible by a suitable choice of $K(X)$ and $G(X)$,
provided that
$r=16\sigma\cs$ remains not too large.
In~\cite{MK} Mizuno and Koyama have studied the case with $\sigma\simeq\nu\ll 1$
focusing their attention on the model~(\ref{gimodel}).
In contrast, the analysis in the present paper
can apply to more general cases with $\sigma\gtrsim 1$.

In the presence of exact shift symmetry 
the exact de Sitter solution is an attractor.
Along this attractor the scalar fluctuations acquire an exactly scale-invariant spectrum.
Making $K$ and/or $G$ weakly dependent on $\phi$,
one obtains a quasi-de Sitter attractor and thereby the spectrum can be tilted.
Though we do not provide corresponding concrete examples here,
more generic, possibly complicated, choices of $K(\phi, X)$ and $G(\phi, X)$
would lead to the interesting situation mentioned above:
$n_s-1\ll 1$ with $\epsilon, s, \delta\gg{\cal O}(n_s-1)$.

\subsubsection{Potential driven G-inflation}

In~\cite{HGI} a novel class of inflation models was proposed
in which the energy density is dominated by $\phi$'s potential
but its dynamics is nontrivial due to the $G\Box\phi$ term.
In particular, it was shown that slow-roll inflation can proceed
even if the potential is too steep to support standard slow-roll inflation.
The model examined in~\cite{HGI} is described by
\begin{eqnarray}
K=X-V(\phi),
\quad
G=-g(\phi)X.
\end{eqnarray}
For $g V_\phi\gg 1$,
the effect of the $G\Box\phi$ term dominates in the slow-roll equation of motion for $\phi$ and
the potential is effectively flattened, leading to
slow-roll G-inflation.
In this regime one finds
\begin{eqnarray}
\sigma\simeq\frac{4}{3}\epsilon \quad{\rm and}\quad
\cs^2\simeq\frac{2}{3}.
\end{eqnarray}
Though $\sigma$ could be free from the slow-roll constraint in principle,
in the present case
it is actually related to $\epsilon$ in a way different from
standard slow-roll inflation.
Since $\cs^2\simeq$ const,
the scale-invariant spectrum requires that $\epsilon\ll 1$, and hence $\sigma\ll 1$.

\section{Bispectrum}

In order to evaluate the bispectrum,
we compute the cubic action for ${\cal R}$
working in the ADM formalism~\cite{Malda, LS, Kachru},
\begin{eqnarray}
\D s^2=-N^2\D t^2+h_{ij}\left(\D x^i+N^i\D t\right)\left(\D x^j+N^j\D t\right),
\end{eqnarray}
where
\begin{eqnarray}
h_{ij}=a^2(t)e^{2{\cal R}}\delta_{ij},
\quad N=1+\alpha_1+\alpha_2+\cdots,\quad N_i=a^2\partial_i\left(\beta_1+\beta_2+\cdots\right)
+ \tilde N_{1i}+\cdots,
\end{eqnarray}
with
$\partial^i\tilde N_{ni}=0$.
Here, $\alpha_n$ and $\beta_n$ are ${\cal O}({\cal R}^n)$.
The fluctuation of the scalar field vanishes in this gauge.
At linear order the above metric reduces to Eq.~(\ref{1st-order-metric}).

As pointed out in~\cite{Malda}, we only need to consider
first-order perturbations in $N$ and $N^i$ to get the cubic action.
(This holds true even in the presence of the $G\Box\phi$ term.)
Therefore, it suffices to use the first-order solution of
the constraint equations, Eqs.~(\ref{cst1}) and~(\ref{cst2}),
supplemented with a vanishing first-order vector perturbation, $\tilde N_{1i}=0$.






We plug the solution for $\alpha_1$ and $\beta_1$ into the action and
expand it to third order in ${\cal R}$. After cumbersome multiple integrations by parts,
one ends up with
\begin{eqnarray}
S_3&=&\int\D t\D^3x\, a^3\biggl[
\frac{{\cal C}_1}{H}\dot{\cal R}^3
+
{\cal C}_2{\cal R}\dot{\cal R}^2
+
\frac{{\cal C}_3}{a^4H^2}\partial^2{\cal R}(\partial{\cal R})^2
+
\frac{{\cal C}_4}{a^2H^2}\dot{\cal R}^2\partial^2{\cal R}
+
{\cal C}_5H {\cal R}^2\dot{\cal R}
\nonumber\\&&\qquad
+
\frac{{\cal C}_6}{a^4H}\partial^2{\cal R}(\partial{\cal R}\cdot
\partial\chi)
+
\frac{{\cal C}_7}{a^4}\partial^2{\cal R}(\partial\chi)^2
+
\frac{{\cal C}_8}{a^2}{\cal R}(\partial {\cal R})^2
+
\frac{{\cal C}_9}{a^2}\dot{\cal R}(\partial{\cal R}\cdot\partial\chi)
+\frac{2}{a^3}f({\cal R})\left.\frac{\delta L}{\delta{\cal R}}\right|_1
\biggr],\label{cubac}
\end{eqnarray}
where $\chi:=\partial^{-2}\Lambda$
with
\begin{eqnarray}
\Lambda:=\frac{a^2}{\Theta^2}X{\cal G}\dot{\cal R}=\frac{a^2\sigma }{\cs^2}\dot{\cal R}.
\end{eqnarray}
The dimensionless coefficients are given by
\begin{eqnarray}
{\cal C}_1&=&
-\frac{H}{\Theta}\frac{\sigma}{\cs^2}\left(1+2\frac{{\cal I}}{{\cal G}}\right)
-2\dot \phi X\left(G_X+XG_{XX}\right)\frac{H\sigma}{\cs^2\Theta^2} +\frac{H^2\sigma }{\cs^4\Theta^2},
\\
{\cal C}_2&=&\frac{\sigma}{\cs^2}\left[3-\frac{H^2}{\cs^2\Theta^2}\left(
3+\epsilon+\frac{2\dot\Theta}{H\Theta}\right)\right],
\\
{\cal C}_3&=&-\frac{H^2\dot\phi XG_X}{\Theta^3},
\\
{\cal C}_4&=&\frac{2H^2\dot\phi X\left(G_X+XG_{XX}\right)}{\Theta^3},
\\
{\cal C}_5&=&\frac{\sigma}{2\cs^2H}\frac{\D}{\D t}\left(\frac{H^2\delta}{\cs^2\Theta^2}\right),
\\
{\cal C}_6&=&\frac{2H\dot\phi XG_X}{\Theta^2},
\\
{\cal C}_7&=&  \frac{\sigma}{4}-\frac{\dot\phi XG_X}{\Theta},
\\
{\cal C}_8&=&-\sigma +\frac{H^2}{\Theta^2}\frac{\sigma}{\cs^2}\left(
1-\epsilon-2s-\frac{2\dot\Theta}{H\Theta}\right),
\\
{\cal C}_9&=&\frac{\sigma}{\cs^2}\left(-\frac{2H}{\Theta}+\frac{\sigma}{2}\right),
\end{eqnarray}
where
\begin{eqnarray}
{\cal I}&:=&XK_{XX}+\frac{2X^2}{3}K_{XXX}
+H\dot\phi G_X+6 X^2G_X^2+5H\dot\phi X G_{XX}
+6X^3G_XG_{XX}+2H\dot\phi X^2 G_{XXX}
\nonumber\\&&
-\frac{2X}{3}\left(2G_{\phi X}+XG_{\phi XX}\right).
\end{eqnarray}
The last term is
the field equation which follows from the quadratic action,
\begin{eqnarray}
\left.\frac{\delta L}{\delta {\cal R}}\right|_1=a\left[\frac{\D\Lambda}{\D t}
+H\Lambda-\sigma\partial^2{\cal R}\right],
\end{eqnarray}
multiplied by
\begin{eqnarray}
f({\cal R})=\frac{H\dot\sigma}{4\cs^2\Theta^2\sigma} {\cal R}^2
+\frac{H}{\cs^2\Theta^2}{\cal R}\dot{\cal R}+
\frac{1}{4a^2\Theta^2}\left[
-(\partial{\cal R})^2+\partial^{-2}\partial^i\partial^j(\partial_i{\cal R}\partial_j{\cal R})
\right]
+\frac{1}{2a^2\Theta}\left[
\partial\chi\cdot\partial{\cal R}-\partial^{-2}\partial^i\partial^j(\partial_i{\cal R}\partial_j\chi)
\right].
\end{eqnarray}
In deriving the above cubic action we have not performed any slow-roll expansion,
so that we have kept full generality up to here.
Taking the limit $G\to 0$, $\Theta\to H$, and $\sigma\to\epsilon$,
we can verify that the above equations reproduce the previous result
derived for generic k-inflation models, ${\cal L}_\phi =K(\phi, X)$~\cite{LS, Kachru}.
In particular, the ${\cal C}_3$, ${\cal C}_4$, and ${\cal C}_6$ terms are absent
in that case. The ${\cal C}_5$ term is clearly a higher order term
so that we will neglect it in the following.

Employing the in-in formalism,
the 3-point function can be computed from the following formula:
\begin{eqnarray}
\langle {\cal R}_{{\mathbf k}_1}{\cal R}_{{\mathbf k}_2}{\cal R}_{{\mathbf k}_3}\rangle
= -i\int^t_{t_0}\D t'\langle\left[
 {\cal R}({\mathbf k}_1, t){\cal R}({\mathbf k}_2, t){\cal R}({\mathbf k}_3, t), H_{\rm int}(t')
\right]\rangle,
\end{eqnarray}
where
$t_0$ is some early time when the fluctuation is well inside the horizon,
$t$ is a time several e-foldings after the horizon exit, and
the interaction Hamiltonian is given by
\begin{eqnarray}
H_{\rm int}(t)=-\int\D^3x\,a^3
\left[\frac{{\cal C}_1}{H}\dot{\cal R}^3+ {\cal C}_2{\cal R}
\dot{\cal R}^2+\cdots\right].
\end{eqnarray}
We use Eqs.~(\ref{def-psi}) and~(\ref{Raa})
to evaluate each contribution, which can be conventionally expressed as
\begin{eqnarray}
\langle {\cal R}_{{\mathbf k}_1}{\cal R}_{{\mathbf k}_2}{\cal R}_{{\mathbf k}_3}\rangle
&=&(2\pi)^7\delta^{(3)}(\mathbf{k}_1+\mathbf{k}_2+\mathbf{k}_3){\cal P}_{\cal R}^2
\frac{{\cal A}}{k_1^3k_2^3k_3^3},
\\
{\cal A}&=&\sum_M{\cal A}_M.
\end{eqnarray}
The power spectrum ${\cal P}_{\cal R}$ here is to be calculated for
the mode with $k_t=k_1+k_2+k_3$.

To proceed, we assume that $\nu =$ const, which holds in a wide class of
G-inflation models as described in Sec.~\ref{models}.
We then immediately see that $\sigma=$ const,
and ${\cal C}_3$, ${\cal C}_6$, and ${\cal C}_7$ are all constant in time as well.
The coefficients are explicitly given by
\begin{eqnarray}
{\cal C}_3= -\frac{(1+\sigma)^2(\sigma-\epsilon)}{(1+\epsilon)^3},
\quad
{\cal C}_6=\frac{2(1+\sigma)(\sigma-\epsilon)}{(1+\epsilon)^2},
\quad
{\cal C}_7=\frac{4\epsilon - \sigma (3-\epsilon)}{4(1+\epsilon)}.
\end{eqnarray}
In order to evaluate the contributions from
the $\dot{\cal R}^3$ (${\cal C}_1$) and $\dot{\cal R}^2\partial^2{\cal R}$ (${\cal C}_4$)
terms, we further
assume that ${\cal I}/{\cal G}$ and $\dot\phi X^2G_{XX}/H$ are of the
form
\begin{eqnarray}
\frac{{\cal I}}{{\cal G}}&=&{\cal J}_1+\frac{{\cal J}_2}{\cs^2},
\\
\frac{\dot\phi X^2G_{XX}}{H}&=&\varrho_1+\frac{\varrho_2}{\cs^2},
\end{eqnarray}
where ${\cal J}_1, {\cal J}_2, \varrho_1$, and $\varrho_2$ are constants.
In kinematically driven
G-inflation~\cite{GI}
we indeed have ${\cal I}/{\cal G}=$ const and $\dot\phi X^2G_{XX}/H=$ const
in the de Sitter limit. In
potential driven G-inflation~\cite{HGI} ${\cal I}/{\cal G}\simeq$ const
and $G_{XX}=0$. Therefore, the assumptions made here are sufficiently
general and reasonable.  It then follows that ${\cal C}_1$ and ${\cal
C}_4$ take the form
\begin{eqnarray}
{\cal C}_1&=&\frac{{\cal D}_1}{\cs^2} + \frac{{\cal E}_1}{\cs^4},
\\
{\cal C}_4&=&{\cal D}_4 + \frac{{\cal E}_4}{\cs^2},
\end{eqnarray}
where ${\cal D}_1, {\cal E}_1, {\cal D}_4$, and ${\cal E}_4$
are constant and are given by
\begin{eqnarray}
 {\cal D}_1 &=& - \frac{\sigma(1+\sigma)}{1+\epsilon}
   \left[ 1+2{\cal J}_1 + 2 \frac{\sigma-\epsilon+(1+\sigma) \varrho_1}{1+\epsilon}
   \right], \\
 {\cal E}_1 &=& - \frac{\sigma(1+\sigma)}{1+\epsilon}
   \left[ 2{\cal J}_2 - \frac{1+\sigma}{1+\epsilon} \left( 1- 2 \varrho_2 \right)
   \right], \\
 {\cal D}_4&=& 2 \frac{(1+\sigma)^3}{(1+\epsilon)^3}
   \left[ \frac{\sigma - \epsilon}{1+\sigma} + \varrho_1 \right], \\
 {\cal E}_4 &=& 2 \frac{(1+\sigma)^3}{(1+\epsilon)^3} \varrho_2.
\end{eqnarray}

Each contribution can now be evaluated as
\begin{eqnarray}
{\cal A}_1&=&\frac{3}{2\sigma}(1-\epsilon-s)\left|\frac{\Gamma(q)}{\Gamma(3/2)}\right|^2
\left(\frac{k_1k_2k_3}{2k_t^3}\right)^{n_s-1}
\left[{\cal D}_1 I_1(n_s-1) +\frac{{\cal E}_1}{\cscr^2}I_1(q')\right],
\\
{\cal A}_2 &= &\frac{1}{4}\left|\frac{\Gamma(q)}{\Gamma(3/2)}\right|^2
\left(\frac{k_1k_2k_3}{2k_t^3}\right)^{n_s-1}\left[3 I_2(n_s-1)-\frac{3-\epsilon}{c_{{\rm s}*}^2}
\left(\frac{1+\sigma}{1+\epsilon}\right)^2I_2(q')\right],
\\
{\cal A}_3 &= &\frac{1}{2}\frac{{\cal C}_3}{\sigma c^2_{{\rm s}*}}(1-\epsilon-s)^2
\left|\frac{\Gamma(q)}{\Gamma(3/2)}\right|^2\left(\frac{k_1k_2k_3}{2k_t^3}\right)^{n_s-1}
I_3(q'),
\\
{\cal A}_4&=&\frac{3}{\sigma}
(1-\epsilon-s)^2
\left|\frac{\Gamma(q)}{\Gamma(3/2)}\right|^2\left(\frac{k_1k_2k_3}{2k_t^3}\right)^{n_s-1}
\left[{\cal D}_4 I_4(n_s-1)+\frac{{\cal E}_4}{\cscr^2}I_4(q')\right],
\\
{\cal A}_6&=&\frac{{\cal C}_6}{8\cscr^2}(1-\epsilon-s)
\left|\frac{\Gamma(q)}{\Gamma(3/2)}\right|^2\left(\frac{k_1k_2k_3}{2k_t^3}\right)^{n_s-1}
I_6(q')
\\
{\cal A}_7&=&\frac{{\cal C}_7}{4}\frac{\sigma}{\cscr^2}
\left|\frac{\Gamma(q)}{\Gamma(3/2)}\right|^2\left(\frac{k_1k_2k_3}{2k_t^3}\right)^{n_s-1}
I_7(q'),
\\
{\cal A}_8&=&\frac{1}{8}\left|\frac{\Gamma(q)}{\Gamma(3/2)}\right|^2
\left(\frac{k_1k_2k_3}{2k_t^3}\right)^{n_s-1}
\left[-I_8(n_s-1)+\frac{1+\epsilon-2s}{c_{{\rm s}*}^2}
\left(\frac{1+\sigma}{1+\epsilon}\right)^2I_8(q')\right],
\\
{\cal A}_9&=&\frac{{\cal C}_{9*}}{8}
\left|\frac{\Gamma(q)}{\Gamma(3/2)}\right|^2\left(\frac{k_1k_2k_3}{2k_t^3}\right)^{n_s-1}
I_9(q'),
\end{eqnarray}
where $\cscr$ and ${\cal C}_{9*}$ are evaluated at sound horizon crossing, $k_t y=-1$, and
\begin{eqnarray}
q':=\frac{s-2\epsilon}{1-\epsilon-s}.
\end{eqnarray}
The $k$-dependent functions $I_M$ are given by
\begin{eqnarray}
I_1(z)&:=&\frac{k_1^2k_2^2k_3^2}{k_t^3} \cos\left(\frac{\pi z}{2}\right)\frac{\Gamma(3+z)}{2},
\\
I_2(z)&:=&\cos\left(\frac{\pi z}{2}\right)\left[
\frac{2+z}{k_t}\sum_{i>j}k_i^2k_j^2-\frac{1+z}{k_t^2}\sum_{i\neq j}k_i^2k_j^3\right]
\Gamma(1+z),
\\
I_3(z)&:=&\frac{({\mathbf k}_1\cdot{\mathbf k}_2)k_3^2}{k_t}
\cos\left(\frac{\pi z}{2}\right)\frac{2+z}{2}
\left\{
\Gamma(1+z)+\Gamma(2+z)\left[
\frac{k_1k_2+k_2k_3+k_3k_1}{k_t^2}+(3+z)\frac{k_1k_2k_3}{k_t^3}\right]\right\}+{\rm sym.},
\\
I_4(z)&:=&\frac{k_1^2k_2^2k_3^2}{k_t^3}\cos\left(\frac{\pi z}{2}\right)\frac{(6+z)\Gamma(3+z)}{12},
\\
I_6(z)&:=&\frac{({\mathbf k}_1\cdot{\mathbf k}_2)k_3^2}{k_t}
\cos\left(\frac{\pi z}{2}\right)
\left[
(3+z)\Gamma(1+z)+(3+z)\Gamma(2+z)\frac{k_3}{k_t}-\Gamma(3+z)\frac{k_3^2}{k_t^2}
\right]
+{\rm sym.},
\\
I_7(z)&:=&\frac{({\mathbf k}_1\cdot{\mathbf k}_2)k_3^2}{k_t}
\cos\left(\frac{\pi z}{2}\right)
\left[
\Gamma(1+z)+\Gamma(2+z)\frac{k_3}{k_t}
\right]
+{\rm sym.},
\\
I_8(z)&:=&\cos\left(\frac{\pi z}{2}\right)\left(\sum_ik_i^2\right)
\left[
\frac{k_t}{1-z}-\frac{1}{k_t}\sum_{i>j}k_ik_j-\frac{1+z}{k_t^2}k_1k_2 k_3
\right]
\Gamma(1+z),
\\
I_9(z)&:=&\frac{({\mathbf k}_1\cdot{\mathbf k}_2)k_3^2}{k_t}
\cos\left(\frac{\pi z}{2}\right)
\left[(3+z)
\Gamma(1+z)-\Gamma(2+z)\frac{k_3}{k_t}
\right]
+{\rm sym.}.
\end{eqnarray}
In computing the above we have used the approximation
\begin{eqnarray}
\psi(k, y) \simeq \sqrt{2}\pi{\cal P}^{1/2}_{{\cal R}}(k_t)k_t^{q-3/2}k^{-q}
\left(1+i ky\right)e^{-iky},\quad |ky| \ll 1.
\end{eqnarray}
One can check that setting $\sigma\to\epsilon$ the above equations
reproduce the result of~\cite{NM}.
The field redefinition ${\cal R}\to{\cal R}_n+f({\cal R}_n)$
gives rise to the non-Gaussian amplitude proportional to $\delta$,
which can be ignored in the present approximation.

The shapes of ${\cal A}_3(1, k_2, k_3)/k_2k_3$ and ${\cal A}_6(1, k_2, k_3)/k_2k_3$
are very similar to the equilateral one, as was already pointed out by~\cite{MK}.
The contributions from $\dot{\cal R}^3$ (${\cal C}_1$) and
$\dot{\cal R}^2\partial^2{\cal R}$
(${\cal C}_4$) give the same momentum dependence because
the two terms are essentially equivalent after using the first-order
equation of motion.

The size of the three-point correlation function is conventionally
parameterized by $f_{\rm NL}$ defined as
\begin{eqnarray}
f_{\rm NL}=30\frac{{\cal A}_{k_1=k_2=k_3}}{k_t^3},
\end{eqnarray}
which can be computed straightforwardly by evaluating the amplitude ${\cal A}$
at $k_1=k_2=k_3=k_t/3$.
The exact expression for $f_{\rm NL}$ is given by
\begin{eqnarray}
f_{\rm NL}&=&30\left|\frac{\Gamma(q)}{\Gamma(3/2)}\right|^2\left(\frac{1}{54}\right)^{n_s-1}
\left\{
\frac{3(1-\epsilon-s)}{2\sigma}\left[{\cal D}_1I_1^{\rm equi}(n_s-1)
+\frac{{\cal E}_1}{\cscr^2}I_1^{\rm equi}(q')\right]\right.
\nonumber\\
&&\qquad\qquad
+\frac{3(1-\epsilon-s)^2}{\sigma}\left[
{\cal D}_4\frac{n_s+5}{6}I_1^{\rm equi}(n_s-1)
+\frac{{\cal E}_4}{\cscr^2}\frac{6+q'}{6}I_1^{\rm equi}(q')
\right]
\nonumber\\
&&\qquad\qquad
+\frac{3}{4}I_2^{\rm equi}(n_s-1)+\frac{1}{4\cscr^2}\left[\frac{3\sigma^2}{8}
-\frac{\sigma-\epsilon}{2(1+\epsilon)}-\frac{1+\sigma}{1+\epsilon}\left(
\sigma+(3-\epsilon)\frac{1+\sigma}{1+\epsilon}\right)\right]I_2^{\rm equi}(q')
\nonumber\\
&&\qquad\qquad
-\frac{(1-\epsilon-s)^2}{2\sigma\cscr^2}\frac{(1+\sigma)^2(\sigma-\epsilon)}{(1+\epsilon)^3}
I_3^{\rm equi}(q')
+\frac{3}{8(n_s-2)}I_6^{\rm equi}(n_s-1)
\nonumber\\
&&\qquad\qquad
\left.
+\frac{1-\epsilon-s}{8\cscr^2}\frac{1+\sigma}{1+\epsilon}
\left(3\frac{1+\sigma}{1+\epsilon}+2\frac{\sigma-\epsilon}{1+\epsilon}\right)I_6^{\rm equi}(q')
\right\},
\end{eqnarray}
where we have defined
$I_M^{\rm equi}(z):=k_t^{-3}I_M(z)|_{k_1=k_2=k_3}$, {\em i.e.},
\begin{eqnarray}
I_1^{\rm equi}(z)&:=&\cos\left(\frac{\pi z}{2}\right)\frac{\Gamma(3+z)}{1458},
\\
I_2^{\rm equi}(z)&:=&\cos\left(\frac{\pi z}{2}\right)\frac{(4+z)\Gamma(1+z)}{81},
\\
I_3^{\rm equi}(z)&:=&\cos\left(\frac{\pi z}{2}\right)\frac{(2+z)(39+13z+z^2)\Gamma(1+z)}{2916},
\\
I_6^{\rm equi}(z)&:=&\cos\left(\frac{\pi z}{2}\right)\frac{(17+9z+z^2)\Gamma(1+z)}{243},
\end{eqnarray}
and used the fact that
$I_4^{\rm equi}(z)=(1+z/6)I_1^{\rm equi}(z)$,
$I_7^{\rm equi}(z)=I_2^{\rm equi}(z)/2$,
$I_8^{\rm equi}(z)=3I_6^{\rm equi}(z)/(1-z)$,
and
$I_9^{\rm equi}(z)=I_2^{\rm equi}(z)$
to shorten the expression.
The above generic formula is involved, but
an order of estimate of $f_{\rm NL}$ is found to be
\begin{eqnarray}
f_{\rm NL}={\cal O}\left(\frac{\tilde\sigma^2}{\cs^2}\right)+
{\cal O}\left(\tilde\sigma^2\frac{XG_{XX}}{G_X}\right)+
{\cal O}\left(\tilde\sigma\frac{{\cal I}}{{\cal G}}\right),
\quad
\tilde\sigma:=\max\{1,\,\sigma\}.\label{fnlestimate}
\end{eqnarray}
This is one of the main results of the present paper.

The sound speed at horizon crossing can be written in terms of $k_t$ as
$\cscr\propto k_t^{s/(1-\epsilon-s)}$.
Under our assumptions we see that $f_{\rm NL}$ can be expressed as
$f_{\rm NL}=f_1+f_2/\cscr^2$,
where $f_1$ and $f_2$ depend on $\epsilon, s, \sigma, {\cal D}_{1,4}$,
and ${\cal E}_{1,4}$,
but are independent of $k_t$.
Therefore, the wavenumber dependence of $f_{\rm NL}$ appears only through $\cscr^2$,
so that the tilt $n_{\rm NG}$ is given by
\begin{eqnarray}
n_{\rm NG}-1=-\frac{f_2\cscr^{-2}}{f_1+f_2\cscr^{-2}}\frac{2s}{1-\epsilon-s}.
\end{eqnarray}
If $\epsilon, s\ll 1$ and
the main contribution to $f_{\rm NL}$ is due to a small sound speed,
then we recover the result of~\cite{Kachru}, $n_{\rm NG}-1\simeq -2s$,
even in the presence of the $G\Box\phi$ term.

\begin{figure}[tb]
  \begin{center}
    \includegraphics[keepaspectratio=true,height=55mm]{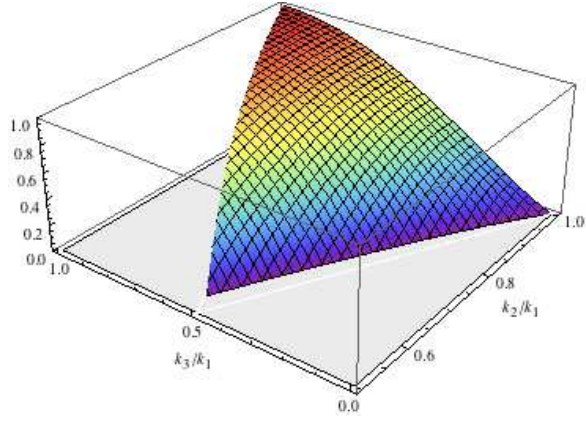}
  \end{center}
  \caption{The non-Gaussian amplitude ${\cal A}(1,k_2/k_1, k_3/k_1)(k_2/k_1)^{-1}(k_3/k_1)^{-1}$
  as a function of $k_2/k_1$ and $k_3/k_1$ for kinematicallly driven G-inflation.
  The amplitude is normalized to unity at an
  equilateral configuration, $k_2/k_1=k_3/k_1=1$. The parameters are given by
  $\sigma = 0.36$, $\cs =0.03$, $\varrho=1$, and ${\cal I}/{\cal G}=1$, so that $r\simeq 0.17$.
  The size of non-Gaussianity is $f_{\rm NL}\simeq 210$.
  }%
  \label{fig:kinetic.eps}
\end{figure}

\begin{figure}[tb]
  \begin{center}
    \includegraphics[keepaspectratio=true,height=55mm]{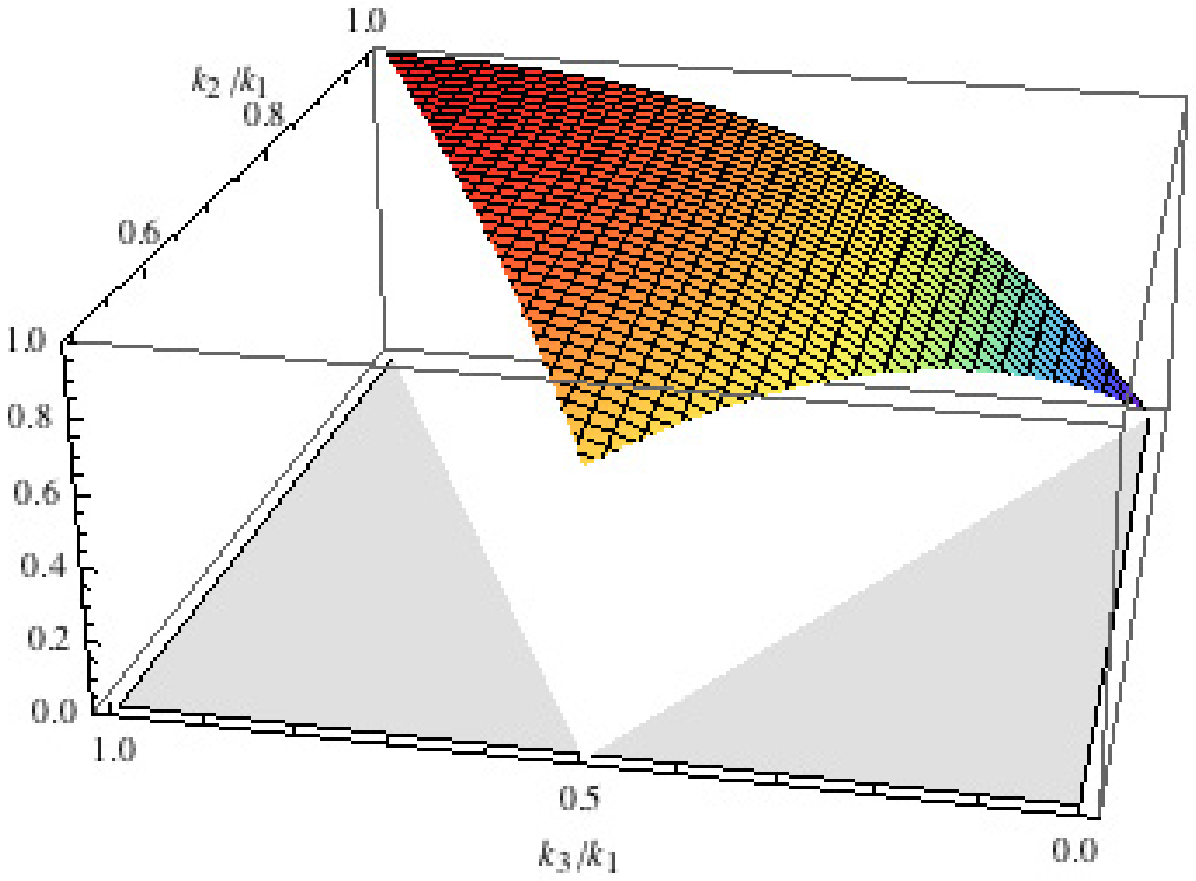}
  \end{center}
  \caption{The non-Gaussian amplitude ${\cal A}(1,k_2/k_1, k_3/k_1)(k_2/k_1)^{-1}(k_3/k_1)^{-1}$
  as a function of $k_2/k_1$ and $k_3/k_1$ for kinematicallly driven G-inflation.
  The amplitude is normalized to unity at an
  equilateral configuration, $k_2/k_1=k_3/k_1=1$. The parameters are given by
  $\sigma = 0.1$, $\cs =0.1$, $\varrho=60$, and ${\cal I}/{\cal G}=1$.
  The size of non-Gaussianity is $f_{\rm NL}\simeq 204$.
  }%
  \label{fig:kin2.eps}
\end{figure}

\begin{figure}[tb]
  \begin{center}
    \includegraphics[keepaspectratio=true,height=55mm]{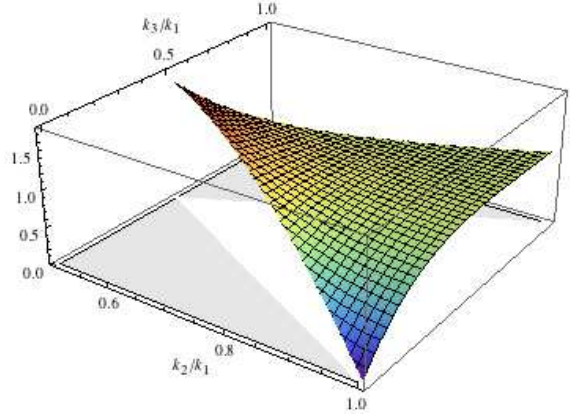}
  \end{center}
  \caption{The non-Gaussian amplitude ${\cal A}(1,k_2/k_1, k_3/k_1)(k_2/k_1)^{-1}(k_3/k_1)^{-1}$
  as a function of $k_2/k_1$ and $k_3/k_1$ for kinematicallly driven G-inflation.
  The amplitude is normalized to unity at an
  equilateral configuration, $k_2/k_1=k_3/k_1=1$. The parameters are given by
  $\sigma = 0.1$, $\cs =0.1$, $\varrho=1$, and ${\cal I}/{\cal G}=300$.
  In this case the shape peaks in the folded configuration $k_1=2k_2=2k_3$.
  }%
  \label{fig:kin3.eps}
\end{figure}

We close this section by illustrating several examples
of non-Gaussian shapes ${\cal A}(1, k_2/k_1, k_3/k_1)(k_2/k_1)^{-1}(k_3/k_1)^{-1}$
in G-inflation.
First, let us consider
the de Sitter limit of shift-symmetric kinematically driven G-inflation.
The functions $K$ and $G$ may be written as
\begin{eqnarray}
K=-X+c_1 X^2+c_2X^3+\cdots,
\quad
G=\frac{X}{M^3}+d_2 X^2+d_3X^3+\cdots,
\end{eqnarray}
where $c_i$, $d_i$, and $M$ are arbitrary in principle.
Given $\nu =\dot\phi XG_X/H$ and $\varrho=\dot\phi X^2G_{XX}/H$,
the former is related to $XK_X/K$ through the background equations,
which in turn fixes the value of $\sigma$.
The latter is related to $\cs^2$, but
since the expression for $\cs^2$ contains both of
the second derivatives $K_{XX}$ and $G_{XX}$, 
$\cs^2$ can be chosen independently of $\varrho$.
Third derivatives $K_{XXX}$ and $G_{XXX}$
appear only in the function ${\cal I}.$
In summary, in the case of kinematically driven G-inflation,
the non-Gaussian amplitude in the de Sitter limit
is completely determined by the four parameters
\begin{eqnarray}
\sigma,\;\; \cs,\;\; \varrho,\;\; \frac{{\cal I}}{{\cal G}}.
\end{eqnarray}
The four parameters can be written in terms of
$c_i$, $d_i$, and $M$, but in practice the expressions are quite involved.
We plot in Figs.~\ref{fig:kinetic.eps}--\ref{fig:kin3.eps} the shapes of non-Gaussianity
for different parameters.

\begin{figure}[tb]
  \begin{center}
    \includegraphics[keepaspectratio=true,height=55mm]{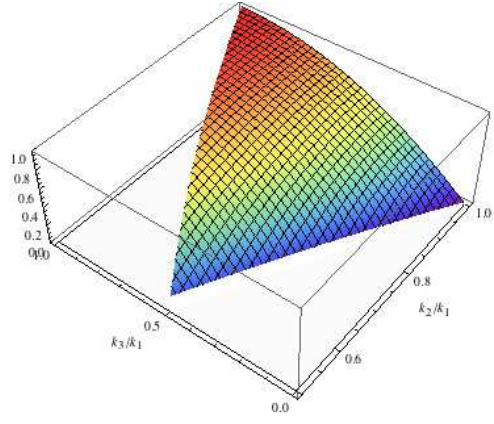}
  \end{center}
  \caption{The non-Gaussian amplitude ${\cal A}(1,k_2/k_1, k_3/k_1)(k_2/k_1)^{-1}(k_3/k_1)^{-1}$
  as a function of $k_2/k_1$ and $k_3/k_1$ for potential driven G-inflation.
  The amplitude is normalized to unity at an
  equilateral configuration, $k_2/k_1=k_3/k_1=1$.
  The size of non-Gaussianity is $f_{\rm NL}=235/3888\simeq 0.06$.
  }%
  \label{fig:potential.eps}
\end{figure}

Another example of the non-Gaussian shapes we explicitly compute
is given by
potential driven G-inflation~\cite{HGI} in the slow-roll approximation.
In this case
we have $\sigma\simeq 4\epsilon/3$, $\cs^2\simeq 2/3$,
$\Theta\simeq H\left(1-\epsilon/3\right)$, and
\begin{eqnarray}
{\cal I}&=&-gH\dot\phi+6g^2X^2+\frac{4}{3}X g_{\phi}
\nonumber\\
&\simeq&\frac{1}{6}{\cal G}.
\end{eqnarray}
The contributions relevant at leading order in slow-roll are:
\begin{eqnarray}
{\cal A}_1\simeq \frac{1}{4}I_1(0),
\quad
{\cal A}_2\simeq -\frac{3}{8}I_2(0),
\quad
{\cal A}_3\simeq -\frac{3}{16}I_3(0),
\quad
{\cal A}_4\simeq \frac{3}{2}I_4(0),
\quad
{\cal A}_8\simeq \frac{1}{16}I_8(0).
\end{eqnarray}
More explicitly, the non-Gaussian amplitude is given by
\begin{eqnarray}
{\cal A}&=&\frac{7}{4}\frac{k_1^2k_2^2k_3^2}{k_t^3}-\frac{3}{8}\left(\frac{2}{k_t}
\sum_{i>j}k_i^2k_j^2-\frac{1}{k_t^2}\sum_{i\neq j}k_i^2k_j^3\right)
\nonumber\\&&
-\frac{3}{16}\left(\frac{1}{2k_t}\sum_ik_i^4-\frac{1}{k_t}\sum_{i>j}k_i^2k_j^2\right)
\left(1+\frac{1}{k_t^2}\sum_{i>j}k_ik_j+3\frac{k_1k_2k_3}{k_t^3}\right)
\nonumber\\&&
+\frac{1}{16}k_t\left(\sum_ik_i^2\right)\left(1-\frac{1}{k_t^2}\sum_{i>j}k_ik_j
-\frac{k_1k_2k_3}{k_t^3}
\right),
\end{eqnarray}
which is plotted in Fig.~\ref{fig:potential.eps}.
Taking the equilateral limit, the size of non-Gaussianity is found to be
$f_{\rm NL}= 235/3888\simeq 0.06$.
The above result is insensitive to the inflaton potential.

\section{Conclusion}

In this paper, we have studied G-inflation, {\em i.e.},
generic single-field inflation
obtained from the Lagrangian~(\ref{Galileon-Lagrangian}).
We have revisited the power spectrum and the spectral index
to clarify how the (approximate) scale-invariance can be achieved
in this class of inflation models,
and determined the possible non-Gaussian amplitude
without assuming slow-roll and the exact scale-invariance.
The nonlinearity parameter $f_{\rm NL}$ in G-inflation
can be summarized schematically as
\begin{eqnarray}
f_{\rm NL}={\cal O}\left(\frac{\tilde\sigma^2}{\cs^2}\right)+
{\cal O}\left(\tilde\sigma^2\frac{XG_{XX}}{G_X}\right)+
{\cal O}\left(\tilde\sigma\frac{{\cal I}}{{\cal G}}\right),
\quad
\tilde\sigma:=\max\{1,\,\sigma\}.
\end{eqnarray}
It should be emphasized that
we have in principle no dynamical constraints that require $\sigma$ to be very small.
If the first term dominates and $\sigma\gtrsim 1$,
then we have
$f_{\rm NL}\sim \sigma^2/\cs^2$.
Therefore, large $f_{\rm NL}$ and large $r=16\sigma\cs$ are compatible.
The situation should be contrasted with the models without the $G\Box\phi$ term,
for which $\sigma=\epsilon$ follows, and hence
large $f_{\rm NL} (\sim 1/\cs^2) $ implies small $r$.

\acknowledgments 
This work was
supported in part by JSPS Grant-in-Aid for Research Activity Start-up
No. 22840011 (T.K.), the Grant-in-Aid for Scientific Research
Nos. 19340054 (J.Y.), 21740187 (M.Y.), and the Grant-in-Aid for
Scientific Research on Innovative Areas No. 21111006 (J.Y.).



\begin{thebibliography}{99}

\bibitem{inflation}
A.~A.~Starobinsky,
  JETP Lett.\  {\bf 30}, 682 (1979)
  [Pisma Zh.\ Eksp.\ Teor.\ Fiz.\  {\bf 30}, 719 (1979)];
A.~H.~Guth,
  Phys.\ Rev.\  D {\bf 23}, 347 (1981);
  K.~Sato,
  Mon.\ Not.\ Roy.\ Astron.\ Soc.\  {\bf 195}, 467 (1981).


\bibitem{wmap}
D.~Larson {\em et al.},
  arXiv:1001.4635 [astro-ph.CO].

  
\bibitem{kinflation}
C.~Armendariz-Picon, T.~Damour and V.~F.~Mukhanov,
  Phys.\ Lett.\  B {\bf 458}, 209 (1999)
  [arXiv:hep-th/9904075].

\bibitem{curvaton}
 A.~D.~Linde, V.~F.~Mukhanov,
  Phys.\ Rev.\  {\bf D56}, 535-539 (1997);
  K.~Enqvist, M.~S.~Sloth,
  Nucl.\ Phys.\  {\bf B626}, 395-409 (2002).
  [hep-ph/0109214];
   D.~H.~Lyth, D.~Wands,
  Phys.\ Lett.\  {\bf B524}, 5-14 (2002).
  [hep-ph/0110002];
  T.~Moroi, T.~Takahashi,
  Phys.\ Lett.\  {\bf B522}, 215-221 (2001).
  [hep-ph/0110096].

\bibitem{susyinf}
K.~A.~Olive,
  Phys.\ Rept.\  {\bf 190}, 307 (1990);
 D.~H.~Lyth and A.~Riotto,
  Phys.\ Rept.\  {\bf 314}, 1 (1999)
  [arXiv:hep-ph/9807278];
D.~H.~Lyth,
  Lect.\ Notes Phys.\  {\bf 738}, 81 (2008)
  [arXiv:hep-th/0702128];
A.~Mazumdar and J.~Rocher,
  Phys.\ Rept.\  {\bf 497}, 85 (2011)
  [arXiv:1001.0993 [hep-ph]];
  M.~Yamaguchi,
  arXiv:1101.2488 [astro-ph.CO].

\bibitem{DBI}
M.~Alishahiha, E.~Silverstein and D.~Tong,
  Phys.\ Rev.\  D {\bf 70}, 123505 (2004)
  [arXiv:hep-th/0404084].


\bibitem{PLANCK}
    [Planck Collaboration],
  arXiv:astro-ph/0604069.


\bibitem{Malda}
J.~M.~Maldacena,
  JHEP {\bf 0305}, 013 (2003)
  [arXiv:astro-ph/0210603].


\bibitem{LS}
D.~Seery and J.~E.~Lidsey,
  JCAP {\bf 0506}, 003 (2005)
  [arXiv:astro-ph/0503692].


\bibitem{Kachru}
X.~Chen, M.~-x.~Huang, S.~Kachru {\it et al.},
  JCAP {\bf 0701}, 002 (2007).
  [hep-th/0605045].

\bibitem{vikman}
  C.~Deffayet, O.~Pujolas, I.~Sawicki {\it et al.},
  JCAP {\bf 1010}, 026 (2010).
  [arXiv:1008.0048 [hep-th]].

\bibitem{GI}
  T.~Kobayashi, M.~Yamaguchi and J.~Yokoyama,
  Phys.\ Rev.\ Lett.\ {\bf 105}, 231302 (2010)
  [arXiv:1008.0603 [hep-th]].
  
  
  \bibitem{G1}
  A.~Nicolis, R.~Rattazzi and E.~Trincherini,
  Phys.\ Rev.\  D {\bf 79}, 064036 (2009)
  [arXiv:0811.2197 [hep-th]].

\bibitem{G2}
  C.~Deffayet, G.~Esposito-Farese and A.~Vikman,
  Phys.\ Rev.\  D {\bf 79}, 084003 (2009)
  [arXiv:0901.1314 [hep-th]];
  C.~Deffayet, S.~Deser and G.~Esposito-Farese,
  Phys.\ Rev.\  D {\bf 80}, 064015 (2009)
  [arXiv:0906.1967 [gr-qc]].






  




\bibitem{gde}
  N.~Chow and J.~Khoury,
  Phys.\ Rev.\  D {\bf 80}, 024037 (2009)
  [arXiv:0905.1325 [hep-th]];
  F.~P.~Silva and K.~Koyama,
  Phys.\ Rev.\  D {\bf 80}, 121301 (2009)
  [arXiv:0909.4538 [astro-ph.CO]];
    T.~Kobayashi, H.~Tashiro and D.~Suzuki,
  Phys.\ Rev.\  D {\bf 81}, 063513 (2010)
  [arXiv:0912.4641 [astro-ph.CO]];
  T.~Kobayashi,
  Phys.\ Rev.\  D {\bf 81}, 103533 (2010)
  [arXiv:1003.3281 [astro-ph.CO]];
  R.~Gannouji and M.~Sami,
  Phys.\ Rev.\  D {\bf 82}, 024011 (2010)
  [arXiv:1004.2808 [gr-qc]];
A.~De Felice, S.~Tsujikawa,
  JCAP {\bf 1007}, 024 (2010).
  [arXiv:1005.0868 [astro-ph.CO]];
A.~De Felice, S.~Mukohyama, S.~Tsujikawa,
  Phys.\ Rev.\  {\bf D82}, 023524 (2010).
  [arXiv:1006.0281 [astro-ph.CO]];
A.~De Felice, S.~Tsujikawa,
  Phys.\ Rev.\ Lett.\  {\bf 105}, 111301 (2010).
  [arXiv:1007.2700 [astro-ph.CO]];
  A.~Ali, R.~Gannouji, M.~Sami,
  Phys.\ Rev.\  {\bf D82}, 103015 (2010).
  [arXiv:1008.1588 [astro-ph.CO]];
  A.~De Felice, S.~Tsujikawa,
  [arXiv:1008.4236 [hep-th]];
  S.~Nesseris, A.~De Felice, S.~Tsujikawa,
  Phys.\ Rev.\  {\bf D82}, 124054 (2010).
  [arXiv:1010.0407 [astro-ph.CO]];
  A.~De Felice, R.~Kase, S.~Tsujikawa,
  Phys.\ Rev.\  {\bf D83}, 043515 (2011).
  [arXiv:1011.6132 [astro-ph.CO]].

\bibitem{HGI}
K.~Kamada, T.~Kobayashi, M.~Yamaguchi and J.~Yokoyama,
  arXiv:1012.4238 [astro-ph.CO].

  
\bibitem{Kimura}
  R.~Kimura, K.~Yamamoto,
  [arXiv:1011.2006 [astro-ph.CO]].
  
\bibitem{ginf1}
C.~Burrage, C.~de Rham, D.~Seery, A.~J.~Tolley,
  JCAP {\bf 1101}, 014 (2011).
  [arXiv:1009.2497 [hep-th]].


\bibitem{ginf2}
  P.~Creminelli, G.~D'Amico, M.~Musso, J.~Norena, E.~Trincherini,
  JCAP {\bf 1102}, 006 (2011).
  [arXiv:1011.3004 [hep-th]].

  
  
\bibitem{eft}
C.~Cheung, P.~Creminelli, A.~L.~Fitzpatrick, J.~Kaplan, L.~Senatore,
  JHEP {\bf 0803}, 014 (2008).
  [arXiv:0709.0293 [hep-th]].

  
\bibitem{NV}
P.~Creminelli, M.~A.~Luty, A.~Nicolis and L.~Senatore,
  JHEP {\bf 0612}, 080 (2006)
  [arXiv:hep-th/0606090].
 
\bibitem{Genesis}
  P.~Creminelli, A.~Nicolis, E.~Trincherini,
  JCAP {\bf 1011}, 021 (2010).
  [arXiv:1007.0027 [hep-th]].



\bibitem{dbigalileon}
  C.~de Rham, A.~J.~Tolley,
  JCAP {\bf 1005}, 015 (2010).
  [arXiv:1003.5917 [hep-th]].




\bibitem{kkreduction}
K.~Van Acoleyen and J.~Van Doorsselaere,
  arXiv:1102.0487 [gr-qc].
  
\bibitem{bstring}
  T.~Kobayashi and T.~Tanaka,
  Phys.\ Rev.\  D {\bf 71}, 084005 (2005)
  [arXiv:gr-qc/0412139].

\bibitem{susy}
J.~Khoury, J.~-L.~Lehners, B.~A.~Ovrut,
  [arXiv:1103.0003 [hep-th]].



\bibitem{MK}
S.~Mizuno and K.~Koyama,
  Phys.\ Rev.\  D {\bf 82}, 103518 (2010)
  [arXiv:1009.0677 [hep-th]].


\bibitem{DeFelice:2011zh}
  A.~De Felice and S.~Tsujikawa,
  arXiv:1103.1172 [astro-ph.CO].

  
\bibitem{naruko}
A.~Naruko, M.~Sasaki,
  [arXiv:1101.3180 [astro-ph.CO]].

\bibitem{per-k-inf}
J.~Garriga and V.~F.~Mukhanov,
  Phys.\ Lett.\  B {\bf 458}, 219 (1999)
  [arXiv:hep-th/9904176].


\bibitem{KP}
J.~Khoury, F.~Piazza,
  JCAP {\bf 0907}, 026 (2009).
  [arXiv:0811.3633 [hep-th]].

\bibitem{NM}
  J.~Noller, J.~Magueijo,
  [arXiv:1102.0275 [astro-ph.CO]].
  
  

\end{thebibliography}
\end{document}